\documentclass[lettersize,journal]{IEEEtran}
\usepackage{amsmath,amsfonts}
\usepackage{algorithmic}
\usepackage{algorithm}
\usepackage{array}

\usepackage{textcomp}
\usepackage{stfloats}
\usepackage{url}
\usepackage{verbatim}
\usepackage{graphicx}
\usepackage{cite}
\usepackage{subfigure}
\usepackage{xcolor}

\usepackage{booktabs}
\begin{document}

\title{Resonant Beam Enabled Passive 3D Positioning}

\author{Yixuan Guo, Mingliang Xiong, Wen Fang, Qingwei Jiang, Mengyuan Xu, Qingwen Liu,~\IEEEmembership{Senior Member,~IEEE,} and Gang Yan,~\IEEEmembership{Member,~IEEE}
\thanks{Y. Guo is with the Shanghai Research Institute for Intelligent Autonomous Systems, Tongji University, Shanghai 201210, China 
(e-mail: guoyixuan@tongji.edu.cn).

M. Xiong is with Hangzhou Institute of Extremely-Weak Magnetic Field Major National Science and Technology Infrastructure, Hangzhou 310052, China (email:xiongml@foxmail.com).

Q. Jiang, M. Xu, and Q. Liu is with the College of Electronics and Information Engineering, Tongji University, Shanghai 201804, China
(e-mail: jiangqw@tongji.edu.cn, xumy@tongji.edu.cn, qliu@tongji.edu.cn).

W. Fang is with the School of Electronic Information and Electrical Engineering, Shanghai Jiao Tong University, Shanghai 200240, China (email: wendyfang@sjtu.edu.cn).

G. Yan is with the School of Physics Science and Engineering, Tongji University, Shanghai 200092, China (e-mail: gyan@tongji.edu.cn).
}

}

\maketitle

\begin{abstract}
With the rapid development of the internet of things (IoT), location-based services are becoming increasingly prominent in various aspects of social life, and accurate location information is crucial. However, RF-based indoor positioning solutions are severely limited in positioning accuracy due to signal transmission losses and directional difficulties, and optical indoor positioning methods require high propagation conditions. To achieve higher accuracy in indoor positioning, we utilize the principle of resonance to design a triangulation-based resonant beam positioning system (TRBPS) in the RF band. The proposed system employs phase-conjugation antenna arrays and resonance mechanism to achieve energy concentration and beam self-alignment, without requiring active signals from the target for positioning and complex beam control algorithms. Numerical evaluations indicate that TRBPS can achieve millimeter-level accuracy within a range of 3.6 m without the need for additional embedded systems.
\end{abstract}


\section{Introduction}
\IEEEPARstart{W}{ith} the rapid development of the intelligent internet of things (IoT), numerous location-based services are emerging in various aspects of social life. These smart devices collect data through sensors to provide users with customized services based on their behavior and location \cite{farahsari2022survey}. Therefore, precise positioning solutions have become an emerging demand in many application scenarios, which improve productivity and quality of life. Examples include logistics and transportation, smart homes, unmanned systems, and indoor positioning.

Currently, mainstream location information is primarily provided by global navigation satellite systems, such as GPS \cite{anjasmara2019accuracy}. However, GPS can only achieve meter-level positioning accuracy, making it more suitable for outdoor applications with lower accuracy requirements. Indoor environments contain many sources of electromagnetic interference, such as \text{WiFi} routers, Bluetooth devices, and other electronic equipment, which can interfere with GPS signals, increase signal delay, and affect positioning accuracy and reliability.

Indoor positioning technologies can be classified into distance-based positioning methods, such as time difference of arrival (TDOA) \cite{jung2011tdoa} and time of flight (TOF)\cite{8288253}, signal strength-based positioning methods, such as received signal strength indicator (RSSI), and angle-based positioning methods, such as direction of arrival (DOA) \cite{obeidat2021review}, \cite{zafari2019survey}. Depending on the electromagnetic wave frequency band, these technologies can be divided into radio frequency-based and light-based positioning technologies.

RF-based positioning technologies mainly include UWB, WiFi, BLE, and Zigbee. UWB technology operates in the 3.1GHz-10.6GHz frequency band and can provide centimeter-level positioning accuracy, which is very high for indoor positioning technologies \cite{mazhar2017precise}. UWB signals have short pulses, resulting in low latency and strong resistance to multipath interference. However, it requires additional hardware on devices, increasing costs. WiFi positioning, also known as 802.11 standard positioning, primarily operates in the 2.4GHz and 5GHz bands. Its most significant advantage is low deployment cost, requiring no special equipment installation \cite{guo2019indoor}. However, WiFi provides poor positioning accuracy and {has high} device power consumption.
BLE is supported by most smart devices today, such as Apple and Samsung, due to its low power consumption and low cost. However, BLE operates in the 2.4GHz to 2.4835GHz frequency range, making it susceptible to WiFi interference. Additionally, it has low accuracy and heavily relies on Bluetooth nodes. Zigbee is a short-range positioning solution mainly operating in the 2.4GHz band, offering low cost, low power consumption, and reliable security \cite{habaebi2014rss}. However, Zigbee has limited positioning distance and high latency, and also it is susceptible to WiFi interference.

Light-based positioning technologies utilize the propagation characteristics of light signals for positioning by emitting, receiving, and analyzing light signals. Visible light communication transmits data by modulating LED light, and receivers achieve positioning by demodulating the signals \cite{du2018demonstration}. LiDAR positioning is achieved by emitting laser pulses and measuring the return time and intensity to construct a 3D image of the environment. Infrared positioning uses infrared emitters and receivers to measure the strength and the arrival time of infrared signals to determine object positions \cite{want1992active}. However, low directionality and power dispersion remain bottlenecks for light-based positioning technologies. Emerging resonant beam systems (RBS) ensure a round-trip light wave between the transmitter and receiver equipped with optical echo reflectors. After multiple iterations, the system reaches stable resonance, demonstrating energy concentration and self-alignment characteristics. This technology can enhance positioning accuracy \cite{liu2016charging}, \cite{liu2022simultaneous}, \cite{fang2021safety}. However, low photoelectric conversion efficiency and high environmental dependency limit its application in indoor positioning.

To compensate for the shortcomings of optical RBS, we design a more adaptable positioning solution by extending RBS to the radio frequency band and redesigning its system structure. Specifically, we equip both the transmitter and receiver with retrodirective antenna (RDA) arrays based on antenna array principles, allowing incident electromagnetic waves to return along the original path. Since the target is passive, we also equip the transmitter with a power amplifier to compensate for unavoidable transmission losses and possible energy consumption of the target. By processing and analyzing the signals received by the transmitter using the MUSIC algorithm and combining triangulation methods, we obtain the 3D coordinates of the passive target.

\begin{figure}
  \centering
        \subfigure[]{
	\includegraphics[width=0.8\linewidth]{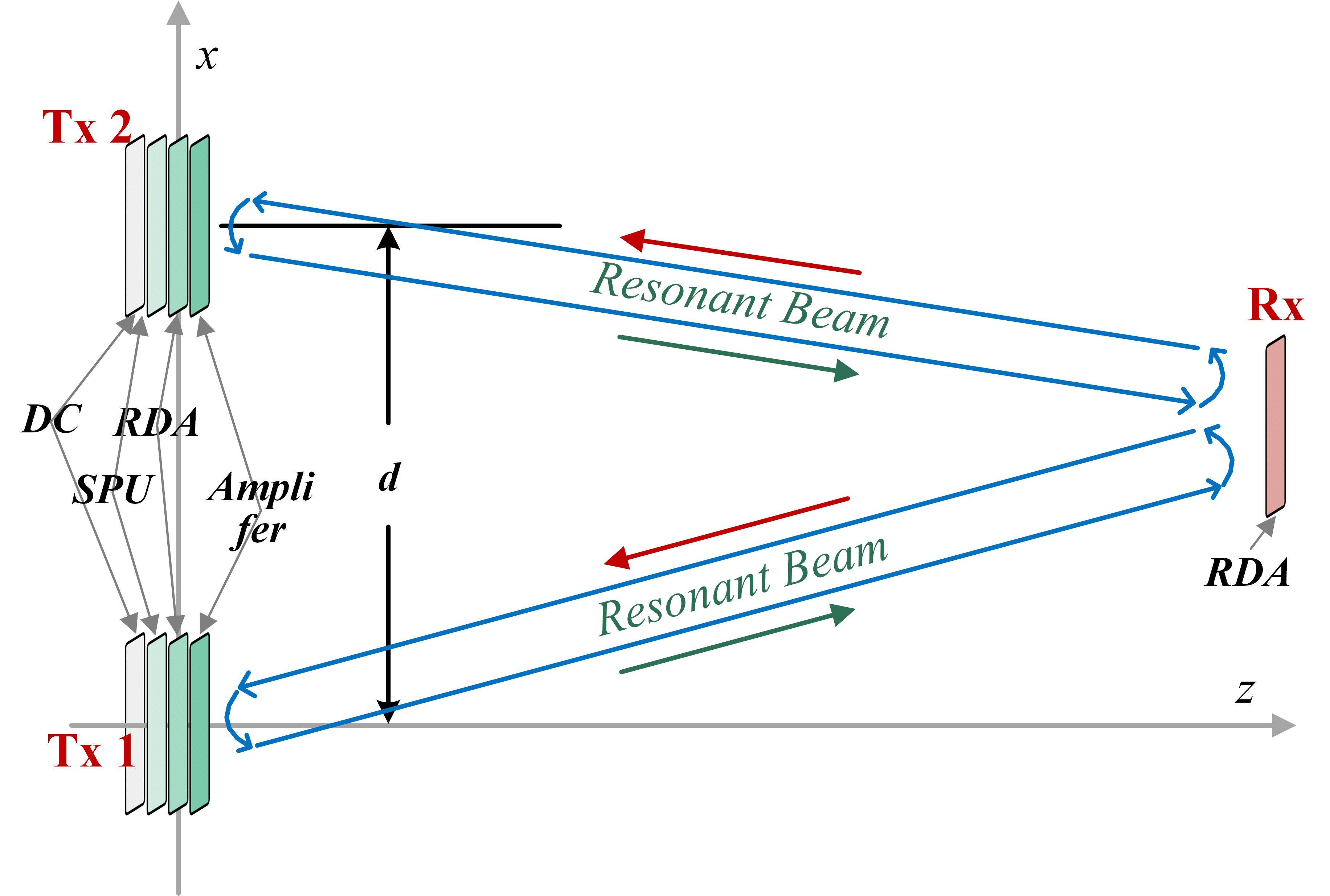}}
 \subfigure[]{
	\includegraphics[width=0.8\linewidth]{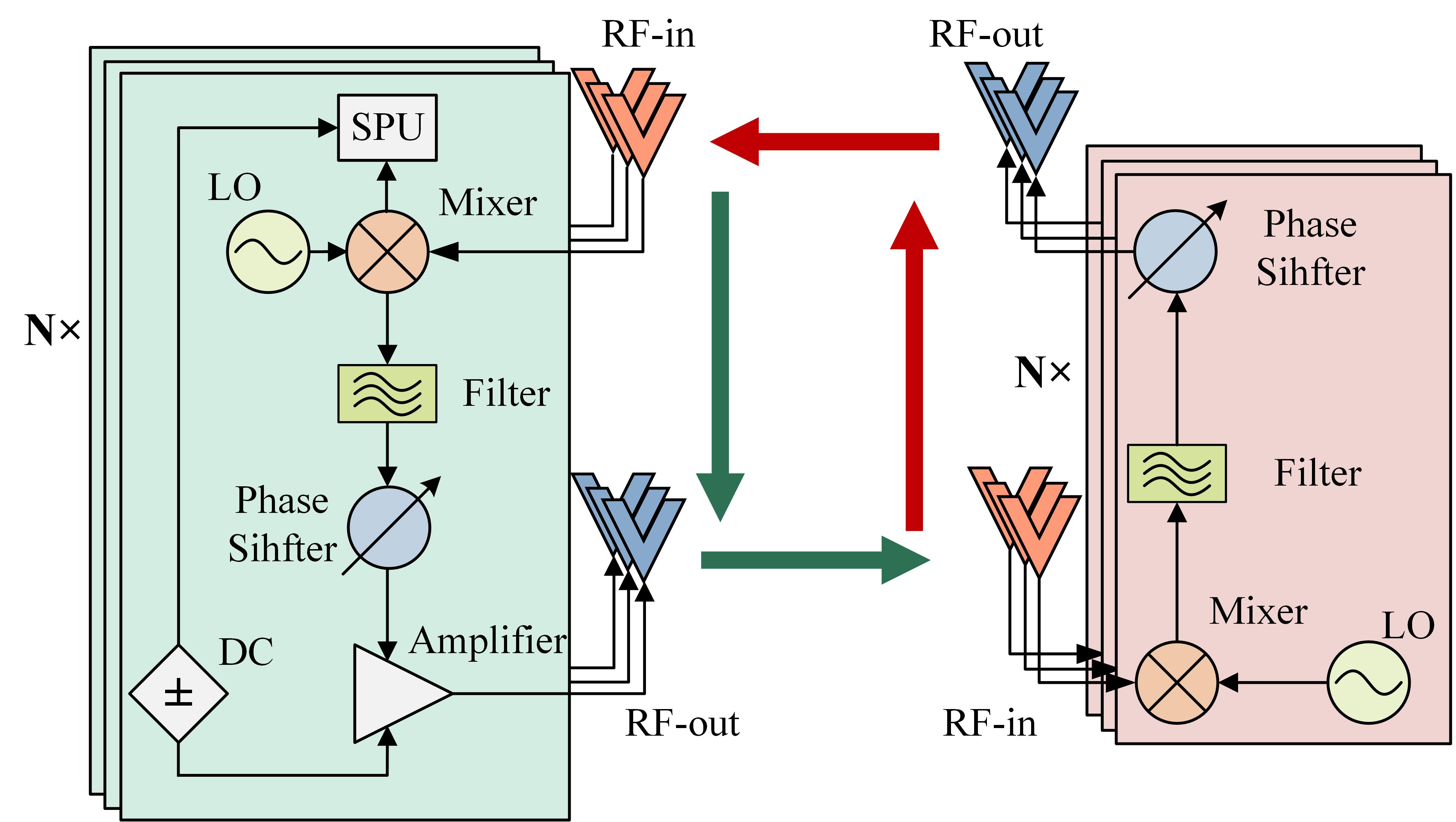}}
\caption{(a) The detailed  structure of transceiver. (b) The structure of the triangulation-based resonant beam positioning system (TRBPS).}
\label{system_design}
\end{figure}

The contributions of this paper are:
\begin{itemize}
    \item [1)] We design a triangulation-based resonant beam positioning system (TRBPS) suitable for the radio frequency band. This system has the characteristics of energy concentration and beam self-alignment, requiring no active signal emission from the target. To our knowledge, this is the first work to propose applying triangulation to RF-RBS for location estimation.

    \item [2)] According to the antenna array principle and the RBS power cycling model, we establish an analytical model of the dual-base resonant beam passive positioning system. Our numerical evaluation shows that the proposed positioning system can directly obtain 3D coordinates without other embedded systems and achieve millimeter-level accuracy.
\end{itemize}

The remainder of this paper is organized as follows. Section II describes the TRBPS architecture, including system structure and positioning principles. Section III establishes the electromagnetic wave cycle model in the resonance system and the analysis model of the passive positioning system of the dual base station resonance wave. Section IV provides a numerical analysis. In Section V, we discuss some open issues. Finally, Section VI provides a conclusion of the paper.

\section{System Overview}
In this section, we first introduce the system design, which mainly includes the system schematic, detailed structure and workflow of each Tx and Rx, and the physical mechanism for the generation of round-trip echoes. Based on this system design, we further introduce that round-trip echoes can form resonance, and based on resonance, 3D positioning of targets can be achieved.

\subsection{System Structure}
As shown in Fig.~\ref{system_design}(a), the TRBP consists of two same transmitters (Tx1 and Tx2) and one receiver (Rx). The two Txs are separated by a distance $d$, and each Tx includes a DC power supply, a signal processing unit (SPU), a power amplifier, and a RDA array. The SPU is mainly used for signal processing and analysis, and it can use spatial signal processing technology to estimate the angle and distance of the received signal source. The power amplifier is used to amplify the output power of the Tx, thereby offsetting the inevitable loss of electromagnetic waves during the round-trip transmission process. The RDA array has a phase conjugation circuit, which allows the signal to return along the original path after phase conjugation processing \cite{leong2003moving}, \cite{dardari2023establishing}.

Compared to Tx, Rx is passive and has a simpler structure, requiring only a RDA array. When the passive target receives the signal, it processes the signal through the phase conjugate circuit, reversing the
signal phase so that it can return along the incident path, thereby achieving signal return. By using reverse arrays at both ends of Tx and Rx, electromagnetic waves can achieve adaptive round-trip echoes.

The specific structure of the transceiver is given in Fig.~\ref{system_design}(b). When Tx receives a signal, it mixes with the local oscillator signal in the mixer to generate an intermediate frequency signal \cite{9921326}. At this point, part of the signal is transmitted to the signal processing unit for target position estimation, and part of the signal is filtered to remove unwanted components, then adjusted for phase by the phase shifter, amplified by the power amplifier, and radiated out. When Rx receives a signal, it mixes with the local oscillator signal in the mixer to generate an intermediate frequency signal, which, after filtering and phase adjustment by the phase shifter, can be directly reflected back.

\begin{figure}[!t]
\centering
\includegraphics[width=0.8\linewidth]{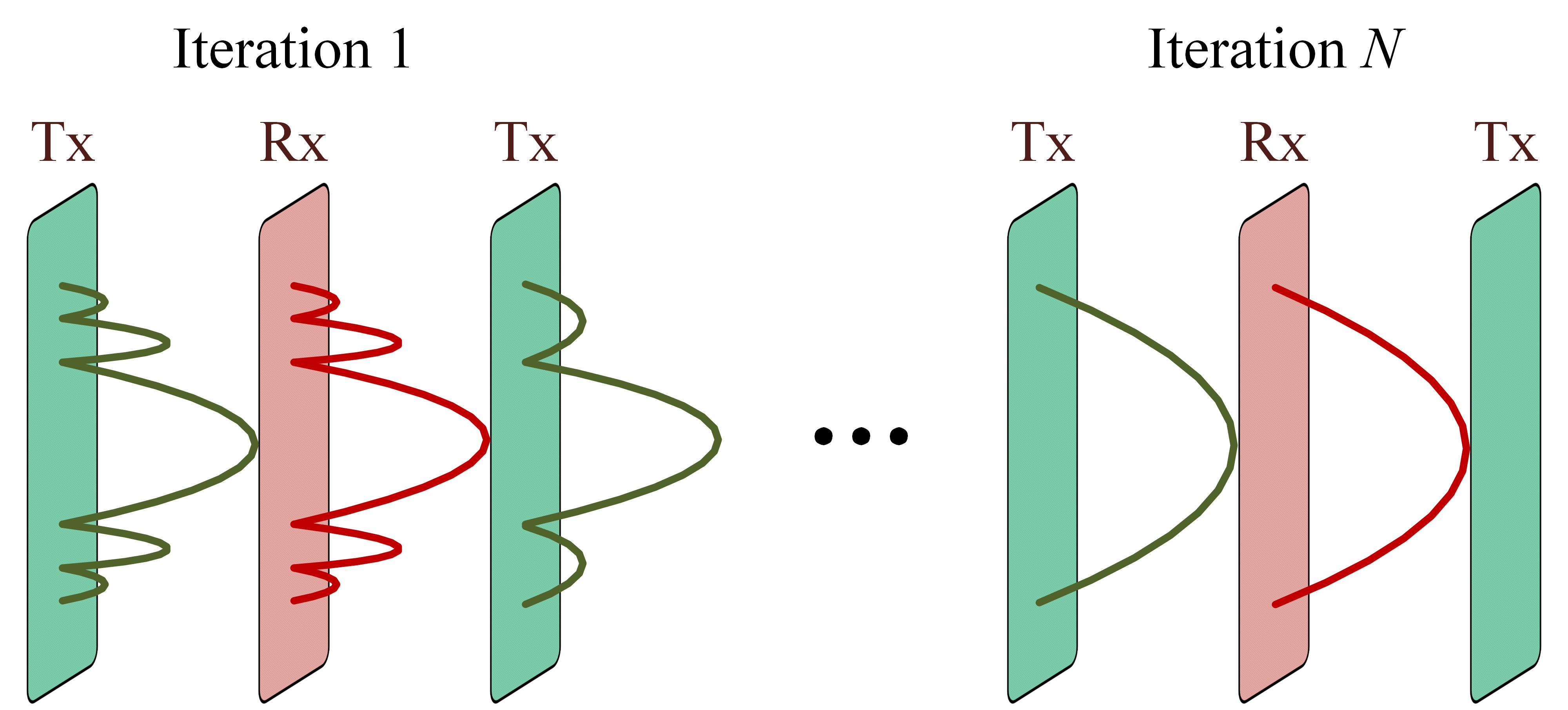}
\caption{The self reproducing mode in TRBPS.}
\label{Reproduce}
\end{figure}

\subsection{Positioning Principle}
In TRBPS, electromagnetic waves radiate back and forth through the RDA arrays at both ends, and cannot form stable oscillations because inevitable losses during transmission weaken the electromagnetic waves strength. Therefore, it is necessary to compensate for spatial transmission losses through Tx's power amplifiers.

When electromagnetic waves are first reflected between two RDA arrays, if the phase and frequency of these waves are the same, they will coherently superimpose to form a standing wave. This standing wave forms a resonance field. After multiple reflections, the strength of the standing wave increases and eventually reaches stable.

On the other hand, electromagnetic waves with different phases will interfere when superimposed, and the phase difference will cause partial wave cancellation. Through multiple reflections and interference, these waves with different phases gradually attenuate and are ultimately eliminated or significantly weakened \cite{aoki2006observation}, \cite{10002391}.

Thus, as shown in Fig.~\ref{Reproduce}, in the continuous oscillation process of electromagnetic waves between each Tx and Rx, the phase distribution gradually self-replicates and reaches a stable state with energy concentration and self-alignment characteristics. Based on these characteristics, we initially transmitted electromagnetic waves simultaneously from two Tx separated by a baseline distance of $d$. The passive Rx's RDA array forms resonant waves between Tx and Rx within a short time after receiving electromagnetic waves from Tx. Each Tx estimates the DOA of the electromagnetic waves returned from Rx using the MUSIC algorithm. Combined with the known baseline distance, the 3D coordinates of the target can be estimated.

\section{ANALYTICAL MODEL}
In this section, we first constructe a power cycling model of the RF resonant beam system based on the principle of resonance. Then, the MUSIC algorithm is used by Tx to estimate the DOA of Rx. Finally, we combine the triangulation method to estimate the 3D coordinates of the Rx.

\begin{figure*}[!t]
\centering
\includegraphics[width=0.8\linewidth]{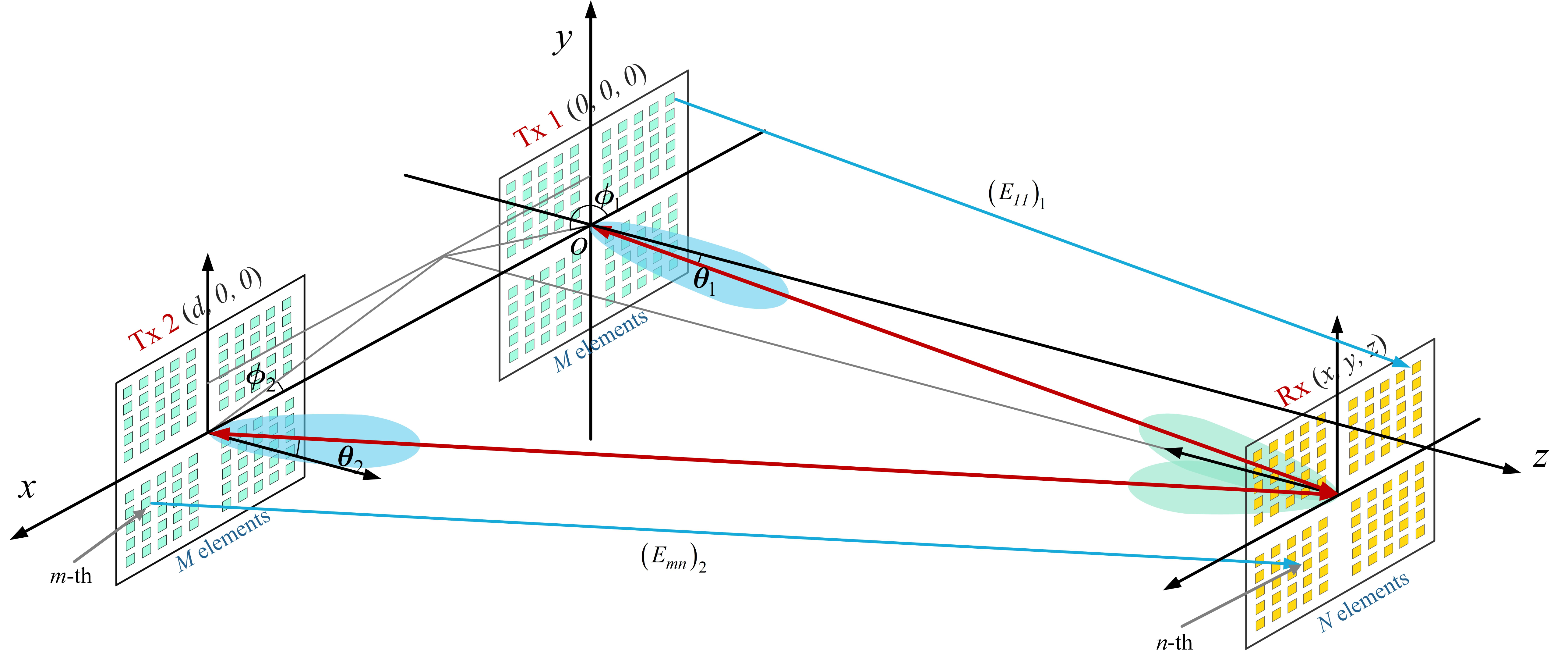}
\caption{The positioning mechanism of TRBPS.}
 \label{system}
\end{figure*}

\subsection{Power Transmission and Reception in Resonance System}

In TRBPS, the power density of electromagnetic waves transmitted back and forth between the Tx array and Rx array, regardless of which end of the RDA array is located, can be expressed by the time-average size of the Poynting vector $\textbf{\textit{S}}$ in free space as
\begin{equation}
    W=|\textbf{\textit{S}}|=\frac{\textbf{\textit{E}}^2}{2\mu},
    \label{e1}
\end{equation}
where the $\mu$ is the wave-impedance and $\textbf{\textit{E}}$ is the electric field. We assume that the Tx array has $M$ antenna elements and the Rx array has $N$ antenna elements \cite{greiner2012classical}. 

As shown in Fig.~\ref{system}, the electric field transmitted from the $m$-th Tx element to the $n$-th Rx element can be expressed as 
\begin{equation}
{\textbf{\textit{E}}_{mn}}=\sqrt{2\mu W_{mn}}e^{-j\varphi_{mn}},
\label{e2}
\end{equation}
where $e^{-j\varphi_{mn}}$ denotes the phase difference caused by distance $l_{mn}$ between the $m$-th Tx element and the $n$-th Rx element. When the $n$-th Rx element initially receives electromagnetic waves from the $m$-th Tx element radiation, the initial phase at Rx can be expressed as
\begin{equation}
    \varphi_{mn}^1=-kl_{mn}+\varphi_0.
\end{equation}

Then, based on the characteristics of the RDA, this element returns the original electromagnetic wave path through a conjugate circuit, and the phase of the electromagnetic wave at this time can be represented as
\begin{equation}
    \varphi_{nm}^1=-\varphi_{mn}^1+\Delta\varphi=kl_{mn}-\varphi_0+\Delta\varphi.
    \label{e4}
\end{equation}

In (\ref{e4}), $\Delta\varphi$ is the phase delay of the conjugate circuit. $W_{mn}$ in (\ref{e2}) is the the power density received from the $m$-th Tx element at the $n$-th Rx element. Under far-field conditions, $W_{mn}$ can be represented by the Friis formula as
\begin{equation}
    W_{mn}=\frac{\lambda^2P_{\mathrm{T}_m}G_{\mathrm{T}_m}G_{\mathrm{R}_n}}{4\pi{l_{mn}^2}(4\pi l_{mn})^2},
    \label{e5}
\end{equation}
where $G_{\mathrm{T}_m}$ and $G_{\mathrm{R}_n}$ are antenna gains of the corresponding elements determined by the directionality of their radiated electromagnetic waves, i.e., $G(\theta, \phi)=G_{max}F(\theta, \phi)$. $P_{\mathrm{T}_m}$ is the transmission power of the $m$-th Tx element, which is initially provided by the DC power supply connected to Tx. In the subsequent iteration process, $P_{\mathrm{T}_m}$ is obtained by receiving and amplifying the electromagnetic waves returned by Rx. By combining (\ref{e2}) and (\ref{e5}), we can get
\begin{equation}
    {\textbf{\textit{E}}_{mn}}=\frac{\lambda e^{-j\varphi_{mn}}}{4\pi {l_{mn}^2}}\sqrt{\frac{\mu}{2\pi}P_{\mathrm{T}_m}G_{\mathrm{T}_m}G_{\mathrm{R}_n}}.
\end{equation}

According to the definition of the power density, the power of the electric field $\textbf{\textit{E}}_{mn}$ can be deduced as
\begin{equation}
    P_{\mathrm{R}_n}=\frac{\lambda^2}{16\pi^2}{\left|\sum_{m=1}^{M}\sqrt{\frac{P_{\mathrm{T}_m}G_{\mathrm{T}_m}G_{\mathrm{R}_n}}{l_{mn}^2}}e^{-j\varphi_{mn}} \right|}^2.
\end{equation}

Furthermore, the total receiving power at the Rx from the Tx can be expressed by
\begin{equation}
        P_{\mathrm{R}}=\frac{\lambda^2}{16\pi^2}\sum_{n=1}^{N}{\left|\sum_{m=1}^{M}\sqrt{\frac{P_{\mathrm{T}_m}G_{\mathrm{T}_m}G_{\mathrm{R}_n}}{l_{mn}^2}}e^{-j\varphi_{mn}} \right|}^2,
\end{equation}
and the transmission efficiency from the Tx to the Rx is
\begin{equation}
    \frac{P_\mathrm{R}}{P_\mathrm{T}}=\frac{\lambda^2}{16\pi^2}\frac{\sum_{n=1}^{N}{\left|\sum_{m=1}^{M}\sqrt{P_{\mathrm{T}_m}G_{\mathrm{T}_m}G_{\mathrm{R}_n}l_{mn}^{-2}}e^{-j\varphi_{mn}} \right|}^2}{\sum_{m=1}^{M}P_{\mathrm{T}_m}}.
    \label{e9}
\end{equation}

When the passive Rx receives power from Tx, it returns electromagnetic waves with a certain power to Tx according to a fixed reflection ratio $\delta$. Then, the Tx amplifies the received electromagnetic waves through an amplifier to offset all possible losses during transmission. Therefore, the output power of the $m$-th Tx antenna in a non-initial state $P_{\mathrm{T}_m}$ can be expressed as
\begin{equation}
    P_{\mathrm{T}_m}=f_{PA}\left(\frac{\lambda^2}{16\pi^2}{\left|\sum_{n=1}^{N}\sqrt{\frac{\delta P_{\mathrm{R}_n}G_{\mathrm{R}_n}G_{\mathrm{T}_m}}{l_{nm}^2}}e^{-j\varphi_{nm}} \right|}^2\right),
\end{equation}
where $f_{PA}$ is the power amplification function which is only determined by the input power. For safety reasons, the maximum output power of each antenna should be limited.

Furthermore, we can observe that (\ref{e9}) primarily calculates the power transmission efficiency based on the gains of Tx and Rx elements ($G_{\mathrm{T}_m}$ and $G_{\mathrm{R}_n}$). These gains are specific to individual antenna elements rather than the entire array. This implies that even in the near-field region of the array antennas (such as the Fresnel zone), the proposed method remains applicable as long as the transmission distance satisfies the far-field condition for individual Tx elements (i.e., $l_{mn}\geq 2D^2/\lambda^2$, where $D$ is the size of a single antenna) \cite{cal_power}.

\subsection{MUSIC Algorithm for DOA Estimation}
The MUSIC algorithm is commonly used to estimate the angle associated with the source of radiation plane waves \cite{gupta2015music}.
Since the total signal power received by Tx is calculated by summing up each antenna element, the distance between a Tx and the Rx only needs to meet the far-field requirements of a single antenna. Therefore, compared to the entire array, our power calculation method is easier to satisfy the far-field assumption.

In RBS, the received signal back-scattered from Rx array can be modelled as
\begin{equation}
\mathbf{X}=\mathbf{A} \mathbf{P} + \mathbf{N},
\end{equation}
where $\mathbf{N} = [N_1, N_2, \ldots, N_M]^T$ is the noise power at each antenna of the Tx, $\mathbf{P} = [P_{{\mathrm{T}_1}}, P_{{\mathrm{T}_2}}, \ldots, P_{{\mathrm{T}_M}}]^T$ is the power of the echo signal received by Tx from passive Rx, and $\mathbf{A} = [\alpha_1, \alpha_2, \ldots, \alpha_K]$ is the manifold matrix, where $\alpha_i$ is the steering vector of the $i$-th source transmit signal, and can be denoted as
\begin{equation}
\alpha_i(\theta, \phi) = \begin{bmatrix}
e^{-j \frac{2\pi}{\lambda} (x_1 \sin \theta \cos \phi + y_1 \sin \theta \sin \phi + z_1 \cos \theta)} \\
e^{-j \frac{2\pi}{\lambda} (x_2 \sin \theta \cos \phi + y_2 \sin \theta \sin \phi + z_2 \cos \theta)} \\
\vdots \\
e^{-j \frac{2\pi}{\lambda} (x_M \sin \theta \cos \phi + y_M \sin \theta \sin \phi + z_M \cos \theta)}
\end{bmatrix}.
\label{e12}
\end{equation}

Each element in (\ref{e12}) represents the phase response of the Rx signal when it reaches each array element from a certain direction. Where $(x_m, y_m, z_m)$ represents the coordinates of the $m$-th Tx element, $\theta$ is the elevation angle, and $\phi$ is the azimuth angle. 

Then, the covariance matrix of the received signal is expressed as
\begin{equation}
    \mathbf{R}=\mathbb{E}[\mathbf{XX}^H].
\end{equation}

The covariance matrix $\mathbf{R}$ is a Hermitian matrix, which can be decomposed into signal subspace and noise subspace through eigenvalue decomposition, as follows

\begin{equation}
    \mathbf{R}=\mathbf{U_X}\mathbf{\Lambda_X} \mathbf{U_X}^H+\mathbf{U_N}\mathbf{\Lambda_N} \mathbf{U_N}^H,
\end{equation}
where $\mathbf{U_X}$ contains the eigenvectors of the signal subspace, corresponding to the largest eigenvalues. $\mathbf{U_N}$ contains the eigenvectors of the noise subspace, corresponding to the smaller eigenvalues. 
$\mathbf{\Lambda}$ is a diagonal matrix, where each diagonal element $\beta_i$ is an eigenvalue of the covariance matrix $\mathbf{R}$. The eigenvalues are typically ordered from largest to smallest
\begin{equation}
    \Lambda=\mathrm{diag}\left(\beta_1, \beta_2, \ldots, \beta_M\right).
\end{equation}

The noise subspace contains the eigenvectors corresponding to the smallest $M$-$K$ eigenvalues. By searching all arrival vectors orthogonal to the noise subspace, the DOA of the passive Rx can be estimated by

\begin{equation}
    (\theta, \phi)_{\mathrm{Tx}}=\mathrm{arg}\min_{\theta, \phi}{\alpha^H(\theta, \phi)\mathbf{U_N}\mathbf{U_N}^H\alpha(\theta, \phi)}.
\end{equation}

\begin{figure}
    \centering
    \includegraphics[width=0.8\linewidth]{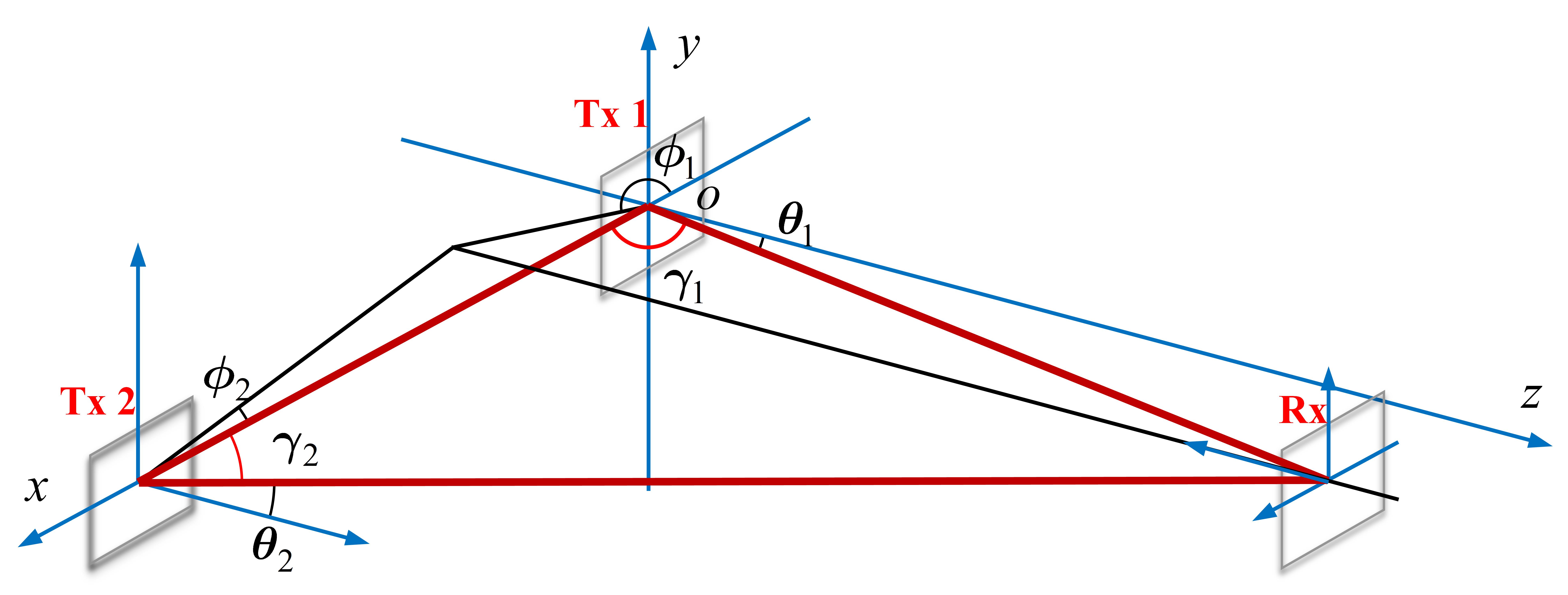}
    \caption{Elevation and azimuth indicator diagram. (a) elevation angle (x-o-z plane), (b) azimuth angle (x-o-y plane).}
    \label{D0A}
\end{figure}

\subsection{Passive Positioning based on Triangulation}
As shown in Fig.~\ref{D0A}, two Txs are used to locate a passive Rx, where the distance between Tx1 and Tx2 is $d$. The azimuth angle $\phi$ is defined as the angle between the projection of the incident electromagnetic wave on the Tx plane (i.e. the x-o-y plane) and the positive x-axis direction, as shown in Fig.~\ref{D0A}(a). The elevation angle $\theta$ is defined as the angle between the projection of the incident electromagnetic wave on the x-o-z plane and the positive z-axis direction, as shown in Fig.~\ref{D0A}(b).

The DOA $(\theta, \phi)$ measured by each Tx based on the MUSIC algorithm can be represented as a unit direction vector. For Tx1, the direction vector can be represented as
\begin{equation}
\mathbf{u}_1 = \begin{bmatrix}
\sin \theta_1 \cos \phi_1 \\
\sin \theta_1 \sin \phi_1 \\
\cos \theta_1
\end{bmatrix}.
\label{e17}
\end{equation}
For Tx2, the direction vector can be represented as
\begin{equation}
\mathbf{u}_2 = \begin{bmatrix}
\sin \theta_2 \cos \phi_2 \\
\sin \theta_2 \sin \phi_2 \\
\cos \theta_2
\end{bmatrix}.
\end{equation}

As shown in Fig~\ref{D0A}, when we form a triangle with Tx1, Tx2, and Rx, the inner angle of the triangle can be expressed as
\begin{equation}
  \begin{cases}
    \gamma_1=\cos^{-1}(\sin\theta_1\cos\phi_1)\\
    \gamma_2=-\cos^{-1}(\sin\theta_2\cos\phi_2)\\
  \end{cases}.
\end{equation}

According to the sine theorem, combined with the known baseline $d$, we can obtain the distance $R_1$ from Rx to Tx1 as
\begin{equation}
R_1=\frac{d\sin(\gamma_2)}{\sin(\gamma_1+\gamma_2)}.
\end{equation}

Assuming the Rx's coordinates are $(x, y, z)$, combined with (\ref{e17}), it can be represented as

\begin{equation}
\begin{bmatrix}
x \\
y \\
z
\end{bmatrix}
=
\begin{bmatrix}
R_1 \sin \theta_1 \cos \phi_1 \\
R_1 \sin \theta_1 \sin \phi_1 \\
R_1 \cos \theta_1
\end{bmatrix}
.
\end{equation}

Finally, we use root mean square error to evaluate the positioning accuracy of the proposed scheme,

\begin{equation}
 \text{RMSE} = \sqrt{\frac{1}{K}\sum_{i=1}^{K}\left(\left(\hat{x_i} - x\right)^2 + \left(\hat{y_i} - y\right)^2 + \left(\hat{z_i} - z\right)^2\right)},
 \label{e22}
\end{equation}
where $K$ is the Monte Carlo simulation number, $\left(\hat{x_i}, \hat{y_i}, \hat{z_i}\right)$ is the estimated coordinate of the Rx, and $(x, y, z)$ is the accurate coordinate of the Rx.

\begin{figure*}
  \centering
        \subfigure[]{
	\includegraphics[width=0.48\linewidth]{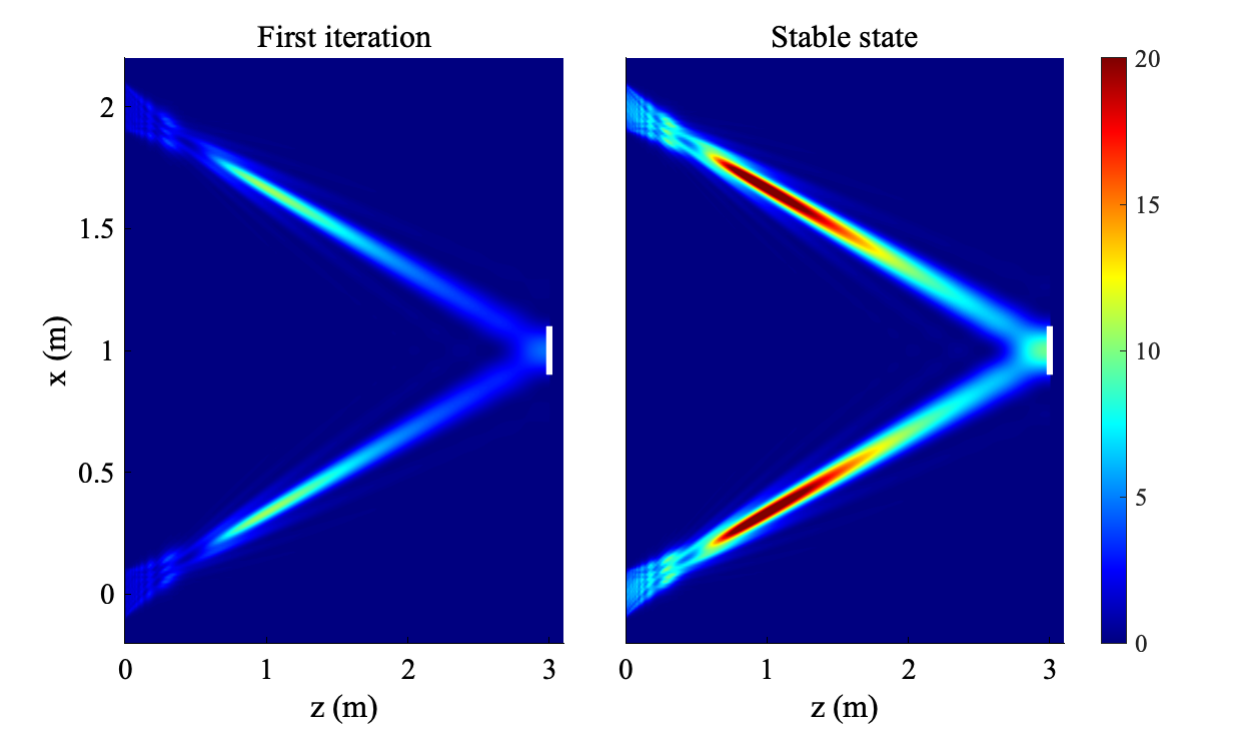}}
 \subfigure[]{
	\includegraphics[width=0.48\linewidth]{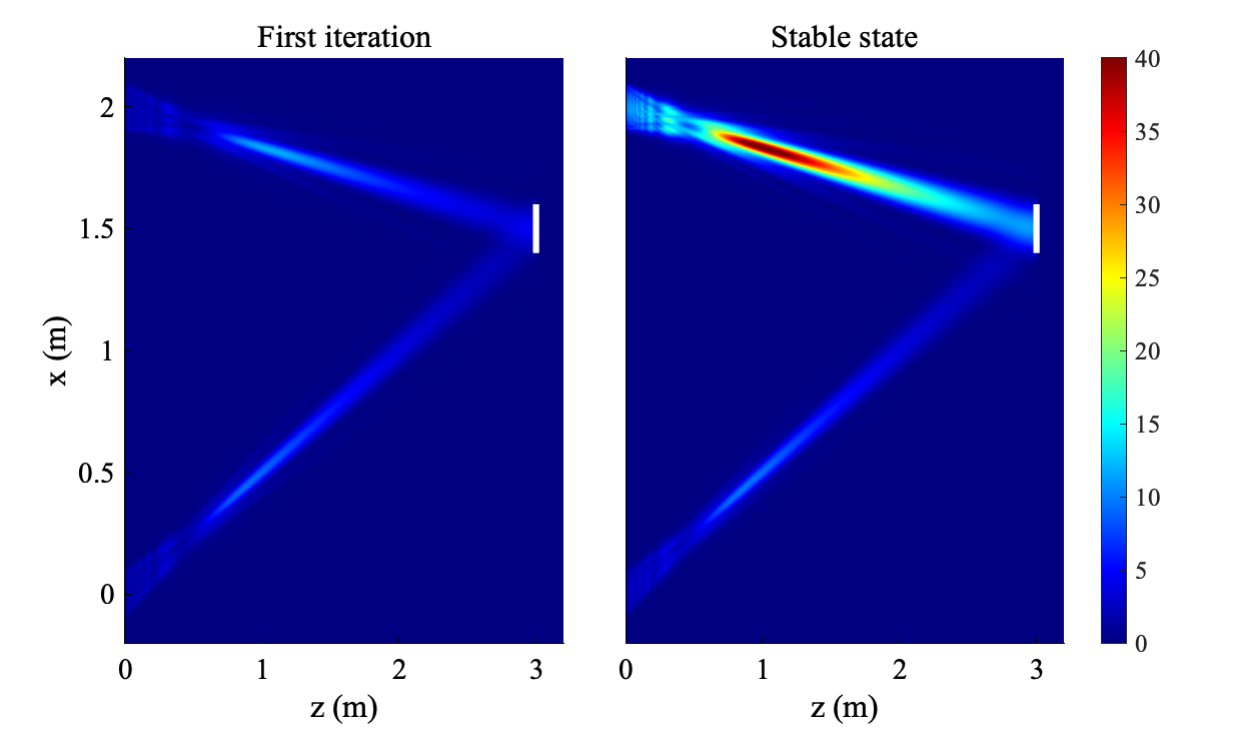}}
\caption{Spatial distribution of power of electromagnetic waves radiated by two Tx and a passive Rx.}
\label{spce_distribution}
\end{figure*}

\section{NUMERICAL ANALYSIS}
In this section, we verify the proposed TRBPS scheme and analyze its performance by conducting numerical simulations. Firstly, we introduce the simulation parameters. Then, we analyze the performance of the proposed system, which mainly involves the electromagnetic wave power cycling model in TRBPS. Finally, we analyze the positioning accuracy of the system, mainly including the estimation accuracy of DOA and 3D coordinates.
\subsection{Parameter Settings}

\begin{table}[h]
    \centering
    \caption{Parameter Setting}
    \begin{tabular}{m{3cm}<{\centering} m{2cm}<{\centering} m{2cm}<{\centering}}
        \toprule 
        \textbf{Parameter} & \textbf{Symbol} & \textbf{Value} \\
        \midrule
        Frequency & $f$ & 30 GHz \\
        Wavelength & $\lambda$ & 1 cm \\
        Interval of elements & $\lambda/2$ & 0.5 cm \\
        Antenna gain \cite{balanis2016antenna}& $G(\theta,\phi)$ & $\leq$4.97 dBi \\     
        Reflection ratio \cite{guo2024resonant}& $\delta$ & 0.004 \\
        Number of elements & $M, N$ & 40$\times$40\\
        Amplifier gain \cite{devices2023}& $G_\mathrm{PA}$ & $\leq$24 dB \\
        Monte Carlo number & $K$ & 100 \\
        Initial input power of Tx & $P_\mathrm{T}^1$ & 1 mW \\
        \bottomrule
    \end{tabular}
    \label{tab:parameter_setting}
\end{table}

Table~\ref{tab:parameter_setting} shows the basic parameters used in our simulation, and unless otherwise specified, these parameters were mostly used in subsequent simulations. In this section, we assume that the carrier frequency of each device is 30GHz, i.e., the wavelength is 1cm. The antenna gain $G(\theta,\phi)$ is the highest in the main direction (i.e. $G(0,0)$), but will not exceed {4.97 dBi}.

\subsection{Performance Analysis of TRBPS}
In Fig.~\ref{spce_distribution}, we simulated the spatial radiation power distribution of TRBPS, where Tx1 is located at $(0, 0, 0)$ and Tx2 is located at $(2, 0, 0)$. In Fig.~\ref{spce_distribution}(a), Rx is located at $(0, 1, 3)$, and the distances to Tx1 and Tx2 are the same, approximately 3.16~m. In Fig.~\ref{spce_distribution}(b), Rx is located at $(0, 1.5, 3)$, and the distances to Tx1 and Tx2 are approximately 3.35~m and 3.04~m, respectively. 

It can be seen that after one round trip, the electromagnetic waves radiated by Tx have initially possessed the characteristics of energy concentration and self-alignment, but with lower power and more sidelobes. After multiple iterations, the electromagnetic wave power radiated by Tx is more concentrated, with stronger directionality, significantly increased power, fewer sidelobes, and stronger directionality, preventing safety hazards caused by radiation to areas outside the target. This is due to the fact that in TRBPS, the phase of the low-power electromagnetic waves initially radiated omnidirectional into space by the Tx's RDA array is uncertain. After receiving some electromagnetic waves radiated to itself, Rx will return them to Tx, which will then be amplified by the power amplifier in Tx to compensate for the inevitable losses during transmission, and continue to return to Rx. During the back and forth oscillation of electromagnetic waves, the distribution of the electromagnetic field continuously replicates and reaches a steady state. The phase of the antennas in the Tx array also gradually stabilizes and aligns, achieving back and forth electromagnetic wave resonance, which can concentrate energy and achieve self-alignment.

Comparing Fig.~\ref{spce_distribution}(a) and (b), it can be seen that the radiation distance and elevation angle have a significant impact on the power transmission. This is mainly because distance and angle constrain the power transmission efficiency. Under the same initial input power conditions, the transmission efficiency of Tx directly affects the power transmission of the system.

\begin{figure}
\centering
\includegraphics[width=\linewidth]{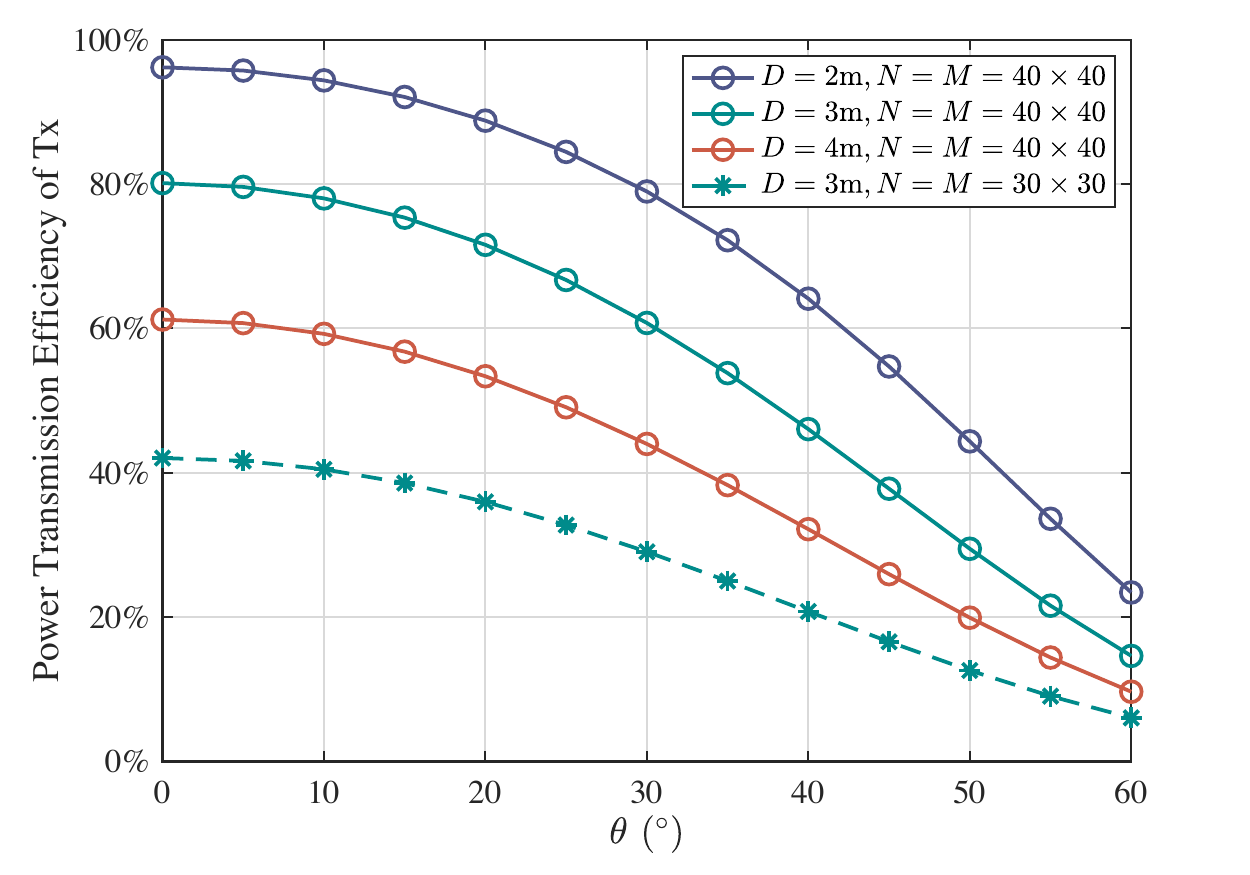}
\caption{Power transmission efficiency versus different elevation angle.}
\label{diff_theta_power}
\end{figure}

In order to conduct a more detailed analysis of the power transmission efficiency of the system, we simulated the transmission efficiency as a function of pitch angle under different array sizes and distance conditions in Fig.~\ref{diff_theta_power}, where the azimuth angle was fixed at $15^\circ$. From the graph, it can be seen that as the pitch angle increases, the transmission efficiency of the system at the same distance continues to decrease. This is due to the decrease in antenna gain caused by the increase in pitch angle, and the decrease in distance also leads to a decrease in transmission efficiency, which can be seen from (\ref{e9}). In addition, a reduction in the number of antenna elements lead to a decrease in the effective receiving area of the antenna, which also results in a decrease in transmission efficiency.

\begin{figure}
\centering
\includegraphics[width=\linewidth]{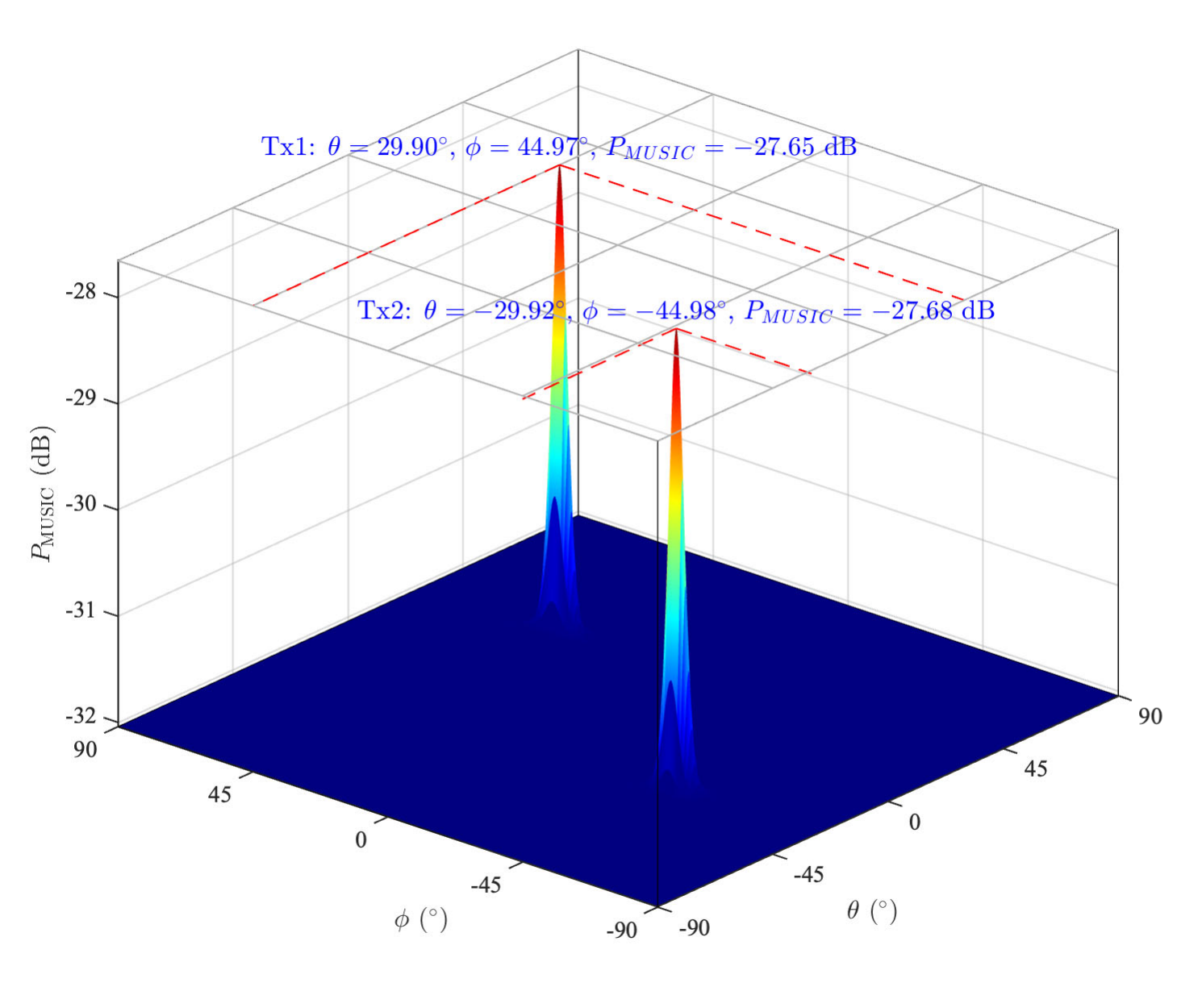}
\caption{3D spectrum obtained by TRPBS through music algorithm.}
\label{3D_MUSIC}
\end{figure}

\subsection{Analysis of 3D Positioning Accuracy}
Figure~\ref{3D_MUSIC} shows the DOA estimation results of two Tx for Rx under the condition of noise power of 0.02 mW using MUSIC algorithm. The theoretical DOA values of Rx radiated electromagnetic waves reaching Tx1 are $\theta_1=30^\circ$  and $\varphi_1=45^\circ$, while the theoretical DOA values of Rx radiated electromagnetic waves reaching Tx2 are $\theta_2=-30^\circ$  and $\varphi_2=-45^\circ$. It can be seen that two spectral peaks obtained through the MUSIC algorithm are sharp and highly accurate, which is also due to the energy concentration and self-alignment characteristics of TRBPS. After multiple iterations, the power received by Tx from Rx gradually increases and reaches a stable state, at which point the SNR reaches its highest.

\begin{figure}
\centering
\includegraphics[width=\linewidth]{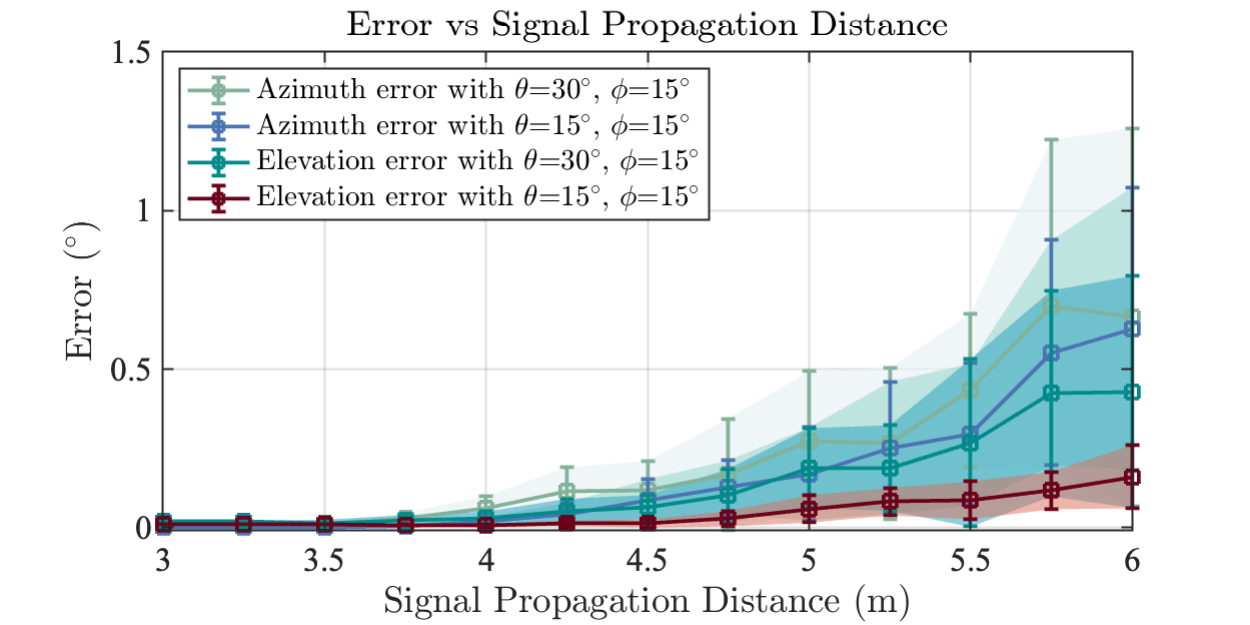}
  \caption{DOA error range versus different distance between a Tx and Rx.}
\label{DOA_error}
\end{figure}

Figure~\ref{DOA_error} shows the impact of distance on DOA estimation. We performed DOA estimation on passive Rx ranging from 3~m to 6~m using the MUSIC algorithm. To ensure the reliability of the results, we conducted 10 simulations for each position and presented the error range. From Fig.~\ref{DOA_error}, it can be seen that the azimuth and elevation errors increase with distance, and the fluctuation range becomes more and more obvious. In addition, the increase in angle will also exacerbate this fluctuation. This conclusion is consistent with Fig.~\ref{diff_theta_power}, mainly because the increase in distance and angle reduces the power transmission efficiency of the resonant system, so the signal power received by Tx is also weak, which affects the signal processing capability and thus reduces the accuracy of DOA estimation.

\begin{figure}
  \centering
        \subfigure[]{
	\includegraphics[width=\linewidth]{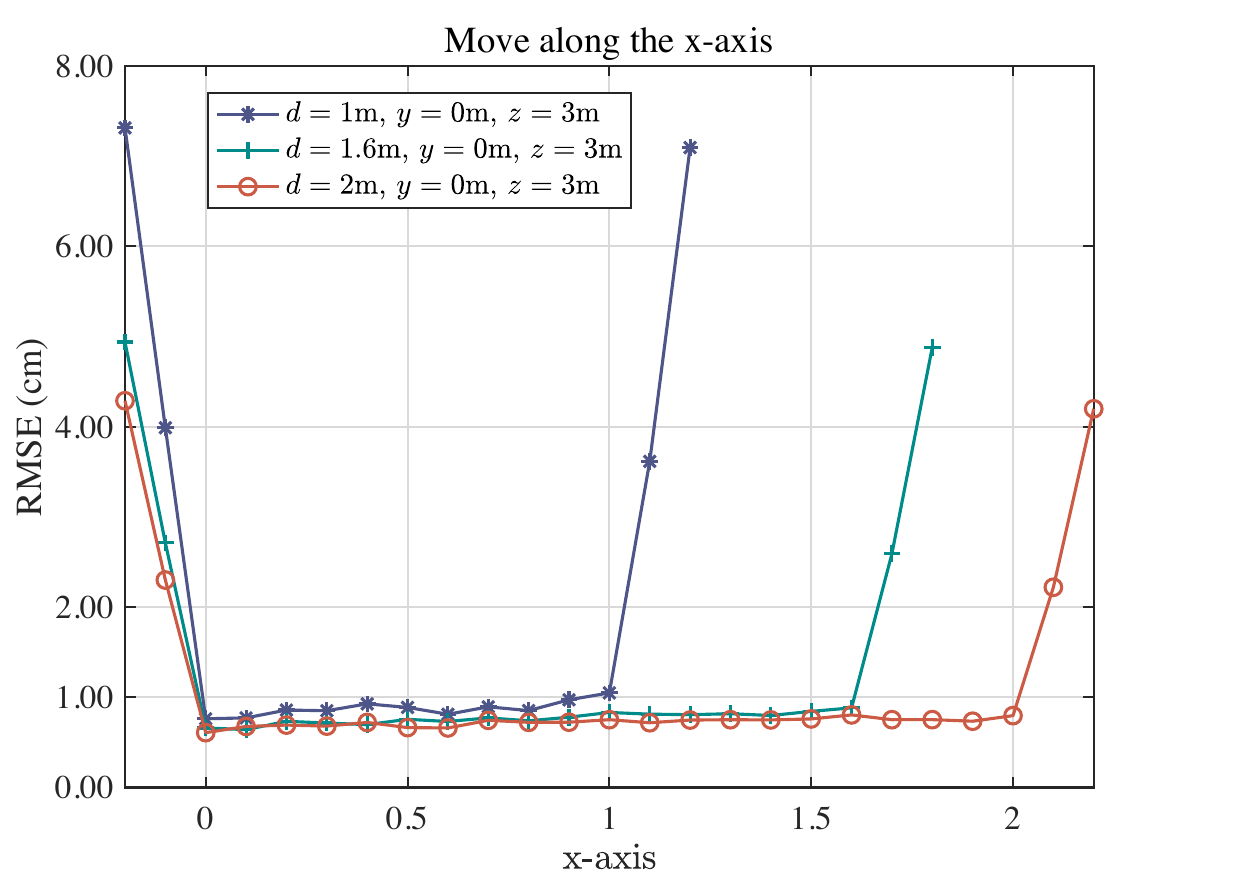}}
 \subfigure[]{
	\includegraphics[width=\linewidth]{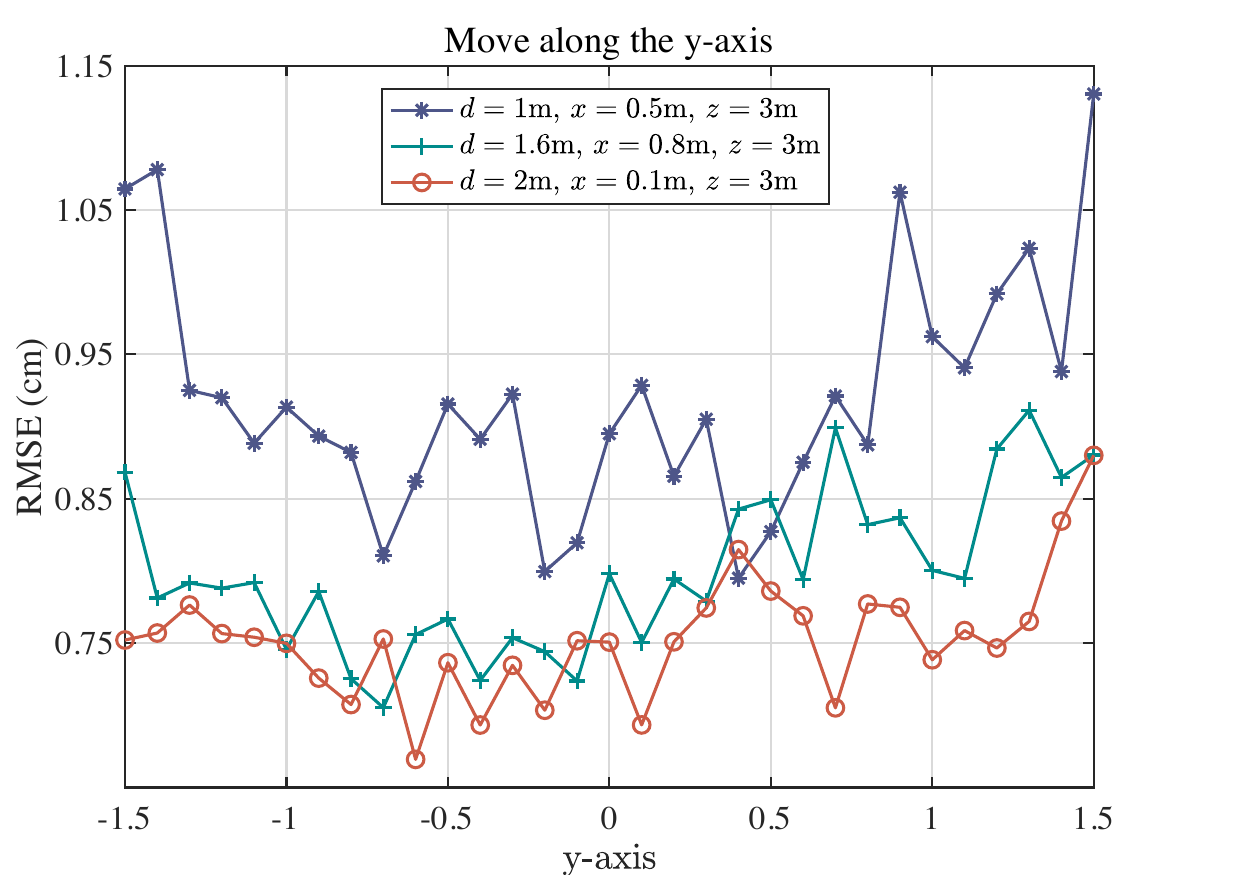}}
  \caption{RMSE of 3D position with different Rx positions.}
  \label{RMSE}
\end{figure}

In order to analyze the positioning performance of the proposed TRBPS more clearly, we estimated the 3D position of Rx at different positions based on triangulation, and quantified the positioning accuracy using the RMSE in (\ref{e22}), as shown in Fig.~\ref{RMSE}. We set the number of Monte Carlo simulations here to 100.

In Fig.~\ref{RMSE}(a), we fix the $y$ and $z$ coordinates of Rx and move it along the $x$-axis, with Rx's coordinates being $(x, 0, 3)$. From the results, it can be seen that when the Rx position changes along the $x$-axis, the positioning errors of different baselines exhibit a symmetrical variation trend with respect to $\frac{d}{2}$, and the maximum error occurs at both ends. This is because in TRBPS, the positioning accuracy is determined by the length of the resonant cavity. In Fig.~\ref{RMSE}(b), we fix the $x$-coordinate and $z$-coordinate of Rx and move it along the $y$-axis, with Rx coordinates being $(\frac{d}{2}, y, 3)$. The results show that as the $y$-coordinate increases, the positioning error first decreases and then increases, reaching its lowest point near $y=0$ m. This is because when Rx moves along the $y$-axis, the distance to Tx1 is the same as the distance to Tx2. When $y=0$ m, the distance to both Tx is the smallest, resulting in the highest accuracy. Overall, the longer the baseline, the slightly higher the positioning accuracy, but the overall positioning accuracy is within 1 cm.
\begin{figure}
  \centering
        \subfigure[]{
	\includegraphics[width=\linewidth]{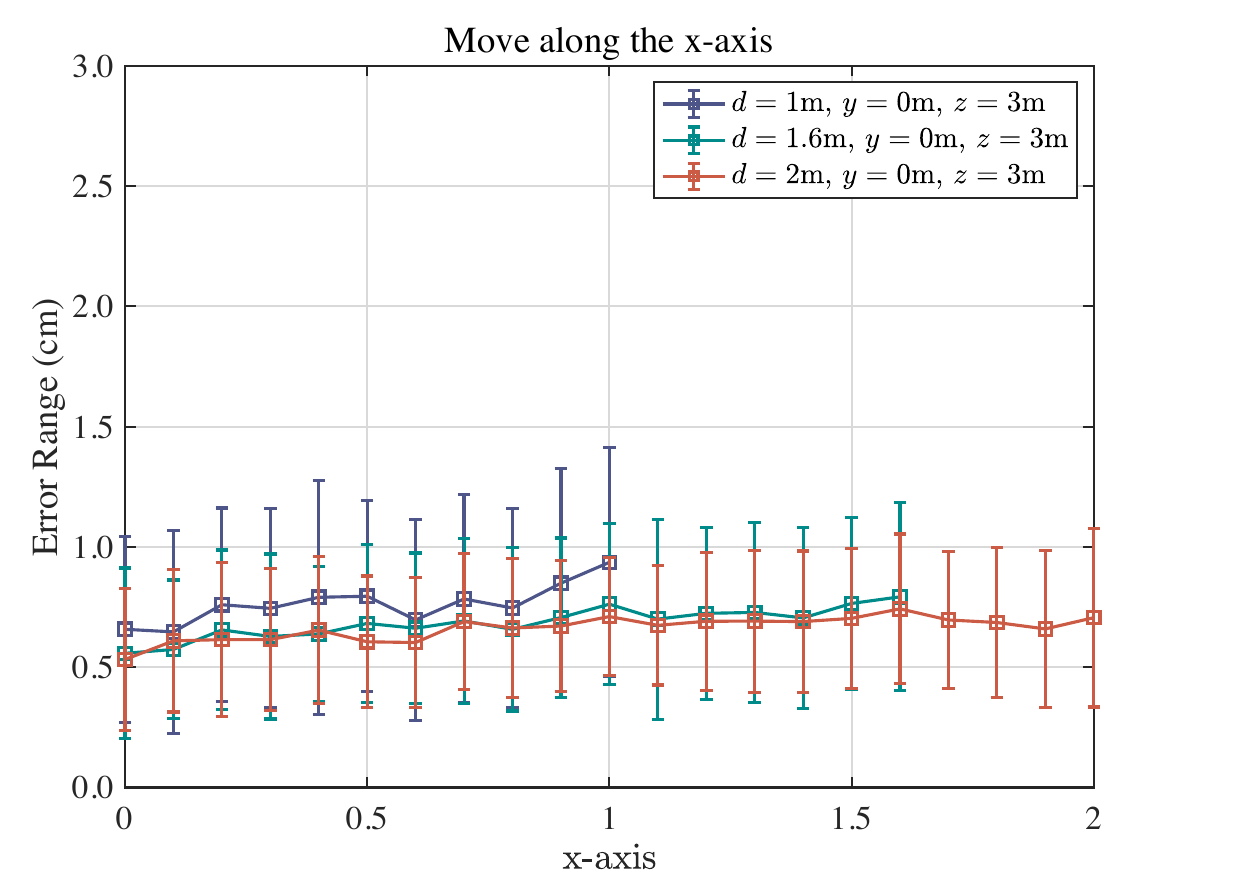}}
	\label{a}
 \subfigure[]{
	\includegraphics[width=\linewidth]{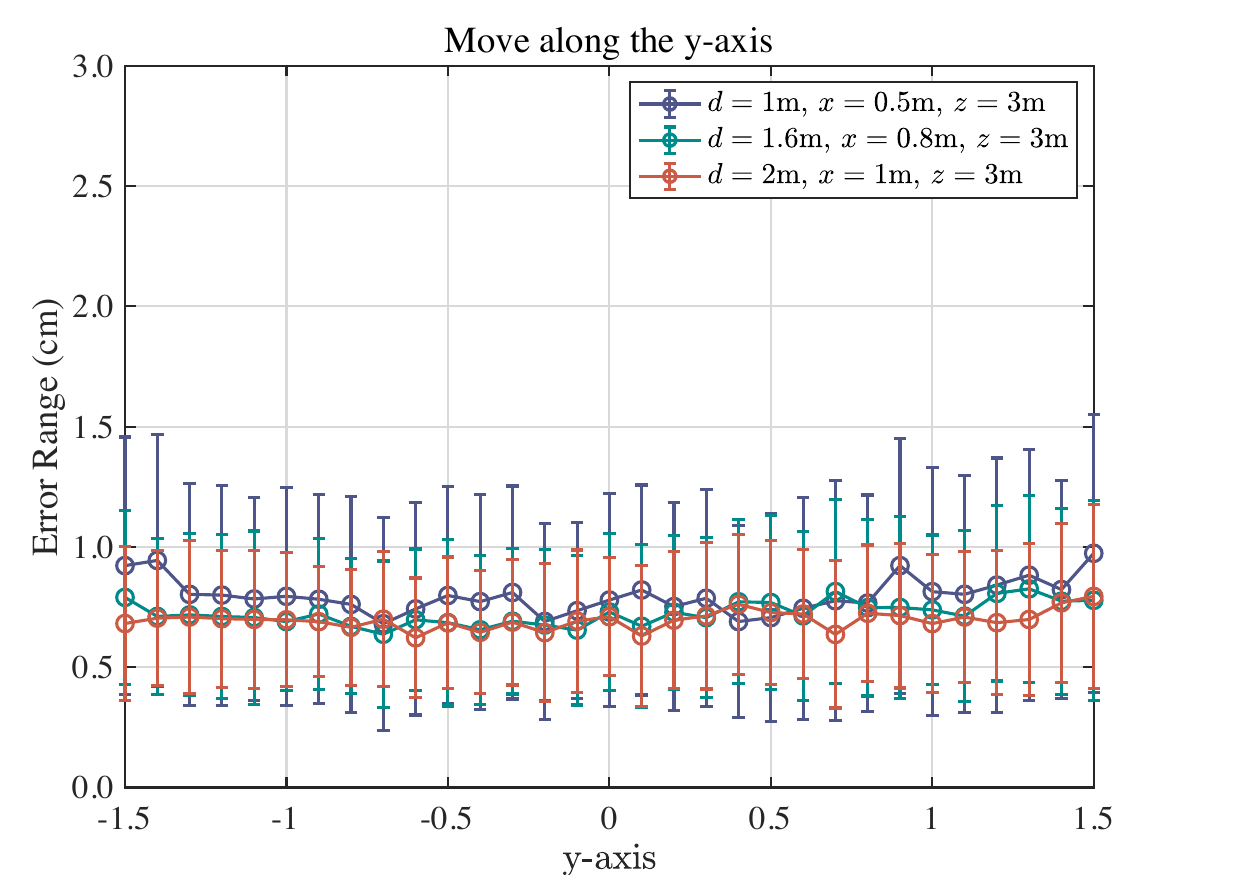}}
        \label{b}
  \caption{Error range of 3D position with different Rx positions.}
  \label{ErrorRange}
\end{figure}

In addition, in order to analyze the positioning performance of TBPRS more clearly, we analyzed the error range of 100 Monte Carlo experiments, as shown in Fig.~\ref{ErrorRange}. Fig.~\ref{ErrorRange}(a) and (b) show the 3D positioning errors of Rx moving from $x=0$~m to $x=d$~m along the $x$-axis and from $y=-1.5$ m to $y=1.5$ m along the $y$-axis at $z=3$ m, respectively. The maximum simulated positioning distance is 3.6 m. From the figure, it can be intuitively seen that the longer the baseline, the higher the positioning accuracy, and the smaller the error fluctuation range, with an error basically not exceeding 1 cm. Therefore, it can be concluded that the proposed TRBPS can achieve millimeter level positioning at a distance of 3.6 m.

\section{Discussion}
In this section, we list two open problems that require further investigation in the future and provide some discussion and ideas for inspiration.

\subsection{Multi-Targets Positioning and Identification}

In theory, each Tx in TRBPS can generate resonant electromagnetic waves with multiple Rx within the radiation range through multiple access techniques such as frequency division multiple access (FDMA), time division multiple access (TDMA), etc. Therefore, the signal processing module of Tx can estimate the angle information of all Rx by receiving signals from different positions Rx, and combine them with  TOF or triangulation to obtain the positions of all targets. 

However, distinguishing each Rx is not an easy task and may require the integration of technologies such as RF fingerprinting \cite{8812713} or RF identification technology \cite{8690994}. Through these technologies, target recognition and access control will be the focus of our next research.

\subsection{Integrated Design of Communication and Positioning}

Integrating communication functions into RBPS not only allows for real-time transmission of positioning data, status information, and other related data but also better serves applications that require timely response and decision-making, such as autonomous driving and dynamic resource scheduling. Additionally, it facilitates easier integration with other systems and devices, supporting various types of data exchange and information sharing, thus making the system more adaptable to different application needs and scenarios.

However, in resonance systems, on one hand, echoes may disrupt resonance stability and cause carrier amplitude fluctuations; on the other hand, echoes, once amplified, become inputs to the modulator, leading to interference in the output signal \cite{8875710}.

Therefore, the critical challenge in the integrated design of communication and localization in resonance systems lies in mitigating the echo effects, which may require redesigning the antenna arrays.

\section{Conclusion}
In this paper, we design a passive positioning system, TRBPS, based on triangulation. Firstly, we construct a positioning system suitable for the RF band based on the resonance principle, featuring energy concentration and self-alignment characteristics. This system allows for stable round-trip propagation of electromagnetic waves between the transmitter and receiver. Subsequently, as the positioning base stations, the two transmitters receive signals reflected back from the passive receiver and obtain the angle information of the Rx relative to each Tx through DOA estimation. Combining this with the known baseline distance between the two Tx, the position of the passive Rx is inferred. Finally, simulations confirm that TRBPS can achieve millimeter-level accuracy within a range of 3.6 m without complex beam control and active signal transmission from target.

\bibliographystyle{IEEEtran}

\bibliography{Mybib}

\newpage

\vfill

\end{document}